\documentclass[aps,pre,twocolumn,groupedaddress,showpacs,showkeys]{revtex4}

\usepackage{graphicx}
\usepackage{dcolumn}
\usepackage{bm}
\usepackage{amssymb}
\usepackage{amsmath}

\usepackage{color}

\newcommand{\alkor}[1]{\textcolor{blue}{#1}}
\renewcommand{\alkor}[1]{#1}

\begin{document}


\title{NEAREST NEIGHBORS, PHASE TUBES AND GENERALIZED SYNCHRONIZATION}



\author{Alexey~A.~Koronovskii}
\author{Olga~I.~Moskalenko}
\author{Alexander~E.~Hramov}
\affiliation{Faculty of Nonlinear
Processes, Saratov State University, Astrakhanskaya, 83, Saratov,
410012, Russia}



\date{\today}

\begin{abstract}
In this paper we report for the first time on the necessity of the refinement of the concept of generalized chaotic synchronization. We show that the state vectors of the
interacting chaotic systems being in the generalized synchronization regime are related with each other by the \emph{functional}, but not the \emph{functional relation} as it was
assumed until now. We propose the phase tube approach explaining the essence of generalized synchronization and allowing the detection and the study of this regime in many relevant physical circumstances. The finding discussed in this Report gives a strong
potential for new applications.
\end{abstract}


\maketitle


Chaotic synchronization is one of the fundamental phenomena, widely
studied recently, having both theoretical and applied
significance~\cite{Boccaletti:2002_ChaosSynchro}. One of the
interesting and intricate types of the synchronous behavior of
unidirectionally coupled chaotic oscillators is generalized
synchronization (GS)~\cite{Rulkov:1996_AuxiliarySystem,
Aeh:2005_GS:ModifiedSystem}. This kind of synchronous behavior is said to mean the presence of a functional relation
between the drive and response oscillator
states~\cite{Rulkov:1995_GeneralSynchro,
Pyragas:1996_WeakAndStrongSynchro} and has been observed in many
systems both
numerically~\cite{Kocarev:1996_GS,Zhigang:2000_GSversusPS,Hramov:2005_GLEsPRE}
and
experimentally~\cite{Rulkov:1996_SynchroCircuits,GS_LightModulator,dmitriev:074101},
with many interesting
features~\cite{Zhigang:2000_GSversusPS,Hramov:2008_INIS_PRE} and
possible
applications~\cite{Terry:GSchaosCom2001,alkor:2010_SecureCommunicationUFNeng}
of this regime being revealed.

The definition of the GS regime generally accepted hitherto is the presence of a functional relation
\begin{equation}
{\mathbf{y}(t)=\mathbf{F}[\mathbf{x}(t)]} \label{eq:FunctRel}
\end{equation}
between the drive $\mathbf{x}(t)$ and response $\mathbf{y}(t)$ oscillator
states~\cite{Rulkov:1995_GeneralSynchro,
Pyragas:1996_WeakAndStrongSynchro}.  Having based on this definition the different techniques for detecting the
presence of GS between chaotic oscillators had been proposed, such as the
nearest neighbor method~\cite{Rulkov:1995_GeneralSynchro,Parlitz:1996_PhaseSynchroExperimental},
the auxiliary
system approach~\cite{Rulkov:1996_AuxiliarySystem} or the conditional Lyapunov
exponent calculation~\cite{Pyragas:1996_WeakAndStrongSynchro}, with the auxiliary system
approach being generally the most easy, clear and powerful tool to study the GS
regime in the model systems, whereas for the analysis of the observed experimental time series the
nearest neighbor method, as a rule, is more applicable~\cite{dmitriev:074101}.

In this Report we report for the first time on the necessity of reconsidering
and refining the existing concept of generalized chaotic synchronization. The
main reason of this refinement is the following. Let
$\mathbf{x}(t_0)=\mathbf{x}_0$ and $\mathbf{y}(t_0)=\mathbf{y}_0$ be the
reference points belonging to the chaotic attractors of the drive and response
oscillators being in the GS regime, respectively. For the neighbor point
$\mathbf{x}(t_i)=\mathbf{x}_i$ of the drive oscillator such that
${||\mathbf{x}_i-\mathbf{x}_0||<\varepsilon}$ its image
$\mathbf{y}(t_i)=\mathbf{y}_i$ in the response system is also close to the
reference point $\mathbf{y}_0$ (see
\cite{Rulkov:1995_GeneralSynchro} for detail), i.e.,
${||\mathbf{y}_i-\mathbf{y}_0||<\delta(\varepsilon)}$. Having linearized
Eq.~(\ref{eq:FunctRel}), one obtains that
\begin{equation}\label{eq:Linearization}
\mathbf{y}_i-\mathbf{y}_0= J\mathbf{F}[\mathbf{x}_0](\mathbf{x}_i-\mathbf{x}_0),
\end{equation}
where $J$ is the Jacobian operator. Since the form of the functional relation
$\mathbf{F}[\cdot]$ can not be found explicitly in most cases,
Eq.~(\ref{eq:Linearization}) may be rewritten in the form
\begin{equation}\label{eq:MatrixForm}
\delta\mathbf{y}_i= \mathbf{A}\delta\mathbf{x}_i,
\end{equation}
where $\mathbf{A}=J\mathbf{F}[\mathbf{x}_0]$ is the unknown matrix and
${\delta\mathbf{x}_i=\mathbf{x}_i-\mathbf{x}_0}$,
${\delta\mathbf{y}_i=\mathbf{y}_i-\mathbf{y}_0}$ are the vectors characterizing
the deviation of the points under consideration $\mathbf{x}_i$, $\mathbf{y}_i$
from the reference points $\mathbf{x}_0$ and $\mathbf{y}_0$, respectively.
Without the lack of generality we shall suppose below the identical dimension
$m$ of the phase space of the drive and response systems.

Although the
coefficients of the matrix $\mathbf{A}$ are unknown, the validity of
Eq.~(\ref{eq:MatrixForm}) may be verified if there are $N>m$ nearest neighbors
$\mathbf{x}_i$ of the reference point $\mathbf{x}_0$ and corresponding them
vectors $\mathbf{y}_i$ of the response system. Having tested the presence of
the generalized synchronization (e.g., with the help of the auxiliary system
approach) we can pick out $m$ nearest neighbors $\mathbf{x}_i$
($i=1,\dots,m$) and corresponding to them vectors $\mathbf{y}_i$ to determine
the coefficients $a_{ij}$ of the matrix $\mathbf{A}$ with the help of
Eq.~(\ref{eq:MatrixForm}). To reduce the influence of the
inaccuracy we have to select such vectors $\mathbf{x}_i$ (and
${\delta\mathbf{x}_i=(\delta x_{i1},\dots,\delta x_{im})^T}$, respectively)
from the whole set of $N$ vectors for which
\begin{equation}\label{eq:BasisCondition}
|\det(\mathbf{X})|=\max,
\end{equation}
where
\begin{equation}\label{eq:XMatrix}
\mathbf{X}=\left(
\begin{array}{cccc}
\delta x_{11} & \delta x_{12} & \ldots & \delta x_{1m} \\
\delta x_{21} & \delta x_{22} & \ldots & \delta x_{2m}\\
\vdots & \vdots & \ddots & \vdots \\
\delta x_{m1} & \delta x_{m2} & \ldots & \delta x_{mm}\\
\end{array}
\right).
\end{equation}
Having determined the matrix $\mathbf{A}$ we can now find the vectors
$\delta\mathbf{z}_i$, (${i=m+1,\dots,N}$) as
\begin{equation}\label{eq:Zvectors}
\delta\mathbf{z}_i= \mathbf{A}\delta\mathbf{x}_i,
\end{equation}
and compare them with the vectors $\delta\mathbf{y}_i$ of the response system
(or compare vectors
${\mathbf{z}_i=\mathbf{y}_0+\delta\mathbf{z}_i}$ with $\mathbf{y}_i$) to
validate the correctness of Eq.~(\ref{eq:MatrixForm}).

Altought, at first sight, it seems that there are no fundamental causes due to
which Eq.~(\ref{eq:MatrixForm}) may fail, in reality Eq.~(\ref{eq:MatrixForm}) is not
correct. To illustrate this fact we have studied numerically the
synchronous behavior of two coupled chaotic R\"ossler oscillators
\begin{equation}
\begin{array}{ll}
\dot x_{d}=-\omega_{d}y_{d}-z_{d},&\dot x_{r}=-\omega_{r}y_{r}-z_{r}
+\varepsilon(x_{d}-x_{r}),\\
\dot y_{d}=\omega_{d}x_{d}+ay_{d},& \dot y_{r}=\omega_{r}x_{r}+ay_{r},\\
\dot z_{d}=p+z_{d}(x_{d}-c), &\dot z_{r}=p+z_{r}(x_{r}-c),\\
\end{array}
\label{eq:Roesslers}
\end{equation}
where $\mathbf{x}=(x_{d},y_{d},z_{d})^T$ [$\mathbf{y}=(x_{r},y_{r},z_{r})^T$] are the
cartesian coordinates of the drive [the response] oscillator, dots
stand for temporal derivatives, and $\varepsilon$ is a parameter
ruling the coupling strength. The other control parameters of Eq.~(\ref{eq:Roesslers}) have been set to $a=0.15$, $p=0.2$, $c=10.0$,
in analogy with our previous studies
~\cite{Aeh:2005_GS:ModifiedSystem,Harmov:2005_GSOnset_EPL}. The
$\omega_r$--parameter (representing the natural frequency of the
response system) has been selected to be $\omega_r=0.95$; the
analogous parameter for the drive system has been fixed to $\omega_d=0.99$.
For such a choice of the parameter values the boundary of
the generalized synchronization regime found with the help of the auxiliary
system approach is around $\varepsilon_{GS}\approx 0.11$.

\begin{figure}[tb]
\centerline{\includegraphics*[scale=0.4]{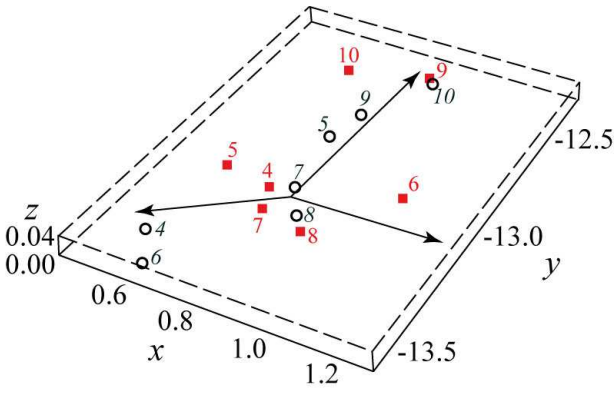}} \caption{(Color
online) The vectors $\mathbf{y}_i$ (\textcolor{red}{$\blacksquare$})
and $\mathbf{z}_i$ (\textcolor{black}{$\circ$}) of the response
R\"ossler system~(\ref{eq:Roesslers}) for $\varepsilon=0.3$. The
numbers $i$ of the vectors are shown by the regular and italic
fonts, respectively} \label{fgr:BadVectors}
\end{figure}

Having chosen the reference  point $\mathbf{x}_0$ of chaotic
attractor of the drive oscillator randomly, one can find its nearest
neighbors $\mathbf{x}_i$ ($i=1,\dots,N$) (and corresponding to them
vectors $\mathbf{y}_i$ of the response system), select (according to
Eqs.~(\ref{eq:BasisCondition}) and (\ref{eq:XMatrix})) the vector
basis $\mathbf{x}_{1-3}$ to determine the matrix $\mathbf{A}$ and
check condition~(\ref{eq:MatrixForm}) with the help of
Eq.~(\ref{eq:Zvectors}) and the rest of vectors $\mathbf{x}_i$,
$\mathbf{y}_i$, ($i=4,\dots,N$).

In Fig.~\ref{fgr:BadVectors} the vectors $\mathbf{z}_i$
($i=4,\dots,10$) obtained with the help of Eq.~(\ref{eq:Zvectors})
as well as the vectors $\mathbf{y}_i$ of the response system are
shown for the coupling strength $\varepsilon=0.3$. The value of the
coupling strength exceeds greatly the threshold $\varepsilon_{GS}$
of the generalized synchronization, the GS regime demonstrate great
stability, and, as a consequence, Eq.~(\ref{eq:MatrixForm}) is
expected to be correct. However, contrary to  expectations, the
vectors $\mathbf{z}_i$ and $\mathbf{y}_i$ differ from each other
sufficiently testifying that Eq.~(\ref{eq:MatrixForm}) fails. As a
matter of fact, the failure of Eq.~(\ref{eq:MatrixForm}) is also
observed for other reference points of the drive R\"ossler
oscillator as well as for other chaotic dynamical systems (e.g.,
Lorenz oscillators). Since Eq.~(\ref{eq:MatrixForm}) is just the
linearization of Eq.~(\ref{eq:FunctRel}), the failure of
Eq.~(\ref{eq:MatrixForm}) is the evidence of the incorrectness of
Eq.~(\ref{eq:FunctRel}) being the main definition of the generalized
synchronization concept. At the same time, plenty of results
obtained hitherto are in the very good agreement with the generally
accepted concept of GS. It means that the concept proposed by
N.~Rulkov et al.~\cite{Rulkov:1995_GeneralSynchro} works in some
circumstances, but, in general, must be refined.

The core idea of this correction is the following. The state of the response system $\mathbf{y}(t)$ depends not only on the state of the drive oscillator $\mathbf{x}(t)$ at the moment of time $t$, but on the history of the evolution of the drive system during time interval $(t-\tau,t]$ as well. Indeed, according to the concept of GS, synchronization means that the response oscillator $\mathbf{y}(t)$ comes to the state defined uniquely by the drive system, with the convergence time $\tau$ being \alkor{connected with} the largest conditional Lyapunov exponent $\lambda_1^r$, i.e. $\tau\sim1/|\lambda_1^r|$. In other words, $\mathbf{F}[\cdot]$ in Eq.~(\ref{eq:FunctRel}) must be considered as \emph{a functional}, but not \emph{a functional relation}. Obviously, in this case Eq.~(\ref{eq:MatrixForm}) obtained under assumptions that $\mathbf{F}[\cdot]$ is the \emph{functional relation} is not satisfied as it has been shown above (see Fig.~\ref{fgr:BadVectors}).

\begin{figure}[tb]
\centerline{\includegraphics*[scale=0.4]{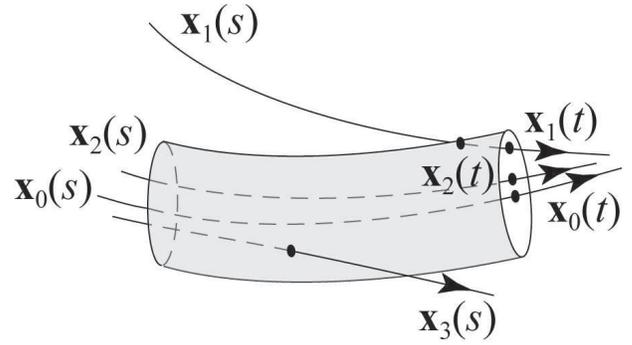}} \caption{The schematic representation of the nearest vectors $\mathbf{x}_i(t)$, the phase trajectories  $\mathbf{x}_i(s)$ and the phase tube $\mathbb{T}_\tau(t)$}
\label{fgr:TubeAndTrajectories}
\end{figure}

Considering $\mathbf{F}[\cdot]$ as the \emph{functional}, one have to replace Eq.~(\ref{eq:Linearization}) by
\begin{equation}\label{eq:FunctionalLinearization}
\delta\mathbf{y}_i(t)= \int\limits_{t-\tau}^t J\mathbf{F}[\mathbf{x}_0(s)]\delta\mathbf{x}_i(s)\,ds.
\end{equation}
Having supposed that the deviation $\delta\mathbf{x}_i(s)$ from the reference trajectory $\mathbf{x}_0(s)$ (${t-\tau < s \leq t}$) is small, in view of the linearity one can write
\begin{equation}\label{eq:Deviations}
\delta\mathbf{x}_{i}(s)=\mathbf{B}(s)\delta\mathbf{x}_{i}(t), \quad
t-\tau < s <t,
\end{equation}
(where $\mathbf{B}(s)$ is the matrix with the time-dependent coefficients) that results in
\begin{equation}\label{eq:FunctionalLinearization2}
\delta\mathbf{y}_i(t)= \int\limits_{t-\tau}^t J\mathbf{F}[\mathbf{x}_0(s)]\mathbf{B}(s)\delta\mathbf{x}_i(t)\,ds.
\end{equation}
and, as a consequence, in
\begin{equation}\label{eq:NearFunctions}
\delta\mathbf{y}_i(t)=\mathbf{C}(t)\delta\mathbf{x}_i(t),
\end{equation}
where $\mathbf{C}(t)$ is the square $(m\times m)$-matrix defined as
\begin{equation}\label{eq:MatrixC}
\mathbf{C}(t)=\int\limits_{t-\tau}^t J\mathbf{F}[\mathbf{x}_i(s)]
\mathbf{B}(s)\,ds.
\end{equation}

\begin{figure}[tb]
\centerline{\includegraphics*[scale=0.4]{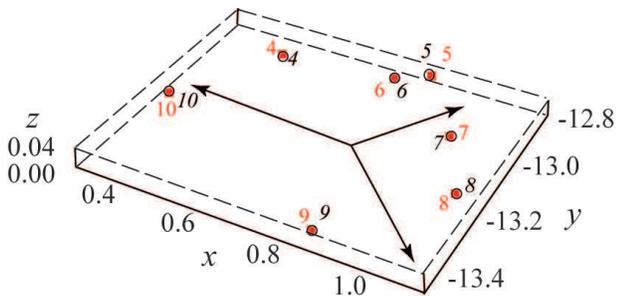}} \caption{(Color
online) The vectors $\mathbf{y}_i$ (\textcolor{red}{$\blacksquare$})
and $\mathbf{z}_i$ (\textcolor{black}{$\circ$}) of the response
R\"ossler system~(\ref{eq:Roesslers}) for $\varepsilon=0.3$, the
length of the phase tube is $\tau=100$. The numbers $i$ of the
vectors are shown by the regular and italic fonts, respectively}
\label{fgr:GoodVectors}
\end{figure}

So, Eq.~(\ref{eq:NearFunctions}) coincides formally with Eq.~(\ref{eq:MatrixForm}) and, therefore, it may be also validated by the calculations of vectors $\mathbf{z}_i$ in the same way as it has been done for Eq.~(\ref{eq:MatrixForm}). At the same time, Eq.~(\ref{eq:MatrixForm}) has been obtained under assumptions that the vectors $\mathbf{x}_0(t)$ and $\mathbf{x}_i(t)$ are close to each other, whereas Eq.~(\ref{eq:NearFunctions}) has been obtained for more stricter restriction requiring the nearness of the \emph{trajectories} $\mathbf{x}_0(s)$ and $\mathbf{x}_i(s)$ during the time interval ${t-\tau<s\leq t}$. Since for the chaotic systems the phase trajectories can converge in one direction of the phase space and diverge in another one, the neighbor vectors $\mathbf{x}_0(t)$ and $\mathbf{x}_i(t)$ may be characterized by the very distinct phase trajectories  $\mathbf{x}_0(s)$ and $\mathbf{x}_i(s)$ for ${t-\tau<s\leq t}$. The schematic representation of such a situation is given in Fig.~\ref{fgr:TubeAndTrajectories}. Although the vectors $\mathbf{x}_1(t)$ and $\mathbf{x}_2(t)$ are close to the reference point $\mathbf{x}_0(t)$, only the vector $\mathbf{x}_2(t)$ obeys Eq.~(\ref{eq:NearFunctions}) due to the nearness of the phase trajectories $\mathbf{x}_0(s)$ and $\mathbf{x}_2(s)$, whereas for the vector $\mathbf{x}_1(t)$ Eq.~(\ref{eq:NearFunctions}) fails, since the phase trajectory $\mathbf{x}_1(s)$ is not close to the reference one $\mathbf{x}_0(s)$ during the whole time interval ${t-\tau<s\leq t}$. Therefore, to verify Eq.~(\ref{eq:NearFunctions}) we have to consider not all vectors $\mathbf{x}_i(t)$ being nearest to the reference point $\mathbf{x}_0(t)$, but only vectors which are characterized by the phase trajectories $\mathbf{x}_i(s)$ being close to the reference one $\mathbf{x}_0(s)$. \alkor{Having based on the idea of phase space strands \cite{Kennel:2002_FalseNneighborsAndFalseStrands,Carroll:2011_AttractorVariationsChaos}}, to eliminate the ineligible vectors (like $\mathbf{x}_1(t)$ in Fig.~\ref{fgr:TubeAndTrajectories}) we introduce into consideration \emph{the phase tube}
\begin{equation}\label{eq:PhaseTube}
\mathbb{T}_\tau(t)=\{\mathbf{x}: \left.\left|x_{0j}(s)-x_j\right|<d_j\right|_{j=1}^m, s\in[t-\tau;t]\}
\end{equation}
and take into account only vectors whose phase trajectories pass through this phase tube (like $\mathbf{x}_2(t)$ in Fig.~\ref{fgr:TubeAndTrajectories}).

The result of this examination for R\"ossler
systems~(\ref{eq:Roesslers}) with the same set of  the control
parameter values and the coupling strength as before is given in
Fig.~\ref{fgr:GoodVectors}, the length of the phase tube is
$\tau=100$. One can see that the calculated vectors
$\mathbf{z}_i(t)$ are in the excellent agreement with the vectors
$\mathbf{y}_i(t)$ of the response R\"ossler system that confirms
both the correctness of Eq.~(\ref{eq:NearFunctions}) and, as a
consequence, the statement that $\mathbf{F}[\cdot]$ is \emph{the
functional}, but not \emph{the functional relation}.

With the increase of the coupling strength between chaotic
oscillators the absolute value of the  largest conditional Lyapunov
exponent $\lambda_1^r$ grows, \alkor{whereas} the time
interval $\tau$ (the length of the phase tube $\mathbb{T}_\tau(t)$)
decreases. Finally, in the lag synchronization (LS) and complete
synchronization (CS) regimes the value of $\tau$ tends to be zero.
Therefore, in the LS and CS regimes Eq.~(\ref{eq:MatrixForm}) is
satisfied for all neighbor vectors $\mathbf{x}_i(t)$ without any
additional requirements concerning the phase trajectory nearness.
\alkor{In other words, the state vectors of any chaotic systems being in the GS regime (but not in the LS or CS regime) are connected with each other with \emph{the functional}, whereas in the LS and CS regimes (which are the strong form of GS) they are related with each other by \emph{the functional relation}.}

\begin{figure}[tb]
\centerline{\includegraphics*[scale=0.45]{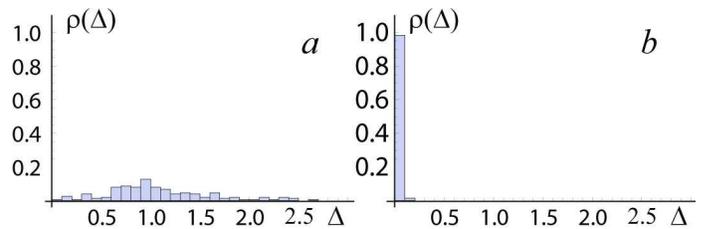}} \caption{\alkor{The histograms of the normalized difference $\Delta=||\delta\mathbf{y}_i(t)-\delta\mathbf{z}_i(t))||/||\delta\mathbf{y}_i(t)||$ for (\textit{a}) the asynchronous dynamics, $\varepsilon=0.06$ and (\textit{b}) the generalized synchronization regime, $\varepsilon=0.3$. The histograms have been obtained for the
response R\"ossler system~(\ref{eq:Roesslers}), the length of the phase tube is $\tau=100$}}
\label{fgr:Histograms}
\end{figure}

Though the phase tube approach has been here applied to the model
systems, we  expect that it can be used in many other relevant
circumstances. \alkor{Since the statistics for the difference between $\delta\mathbf{y}_i(t)$ and $\delta\mathbf{z}_i(t)$ vectors are radically different for the synchronous and asynchronous motion (see Fig.~\ref{fgr:Histograms})}, the important feature of this approach is the
possibility to consider the relation between
vectors~(\ref{eq:NearFunctions}) for the analysis of the registered
experimental data (vector or scalar, using the Takens approach~\cite{Takens:1981}) when the other classical methods of GS detection are inaccurate or unapplicable.
Moreover, the proposed approach may be used as the method to detect the GS
regime, including the case when the chaotic oscillators are coupled
mutually, since all arguments given above are also applicable for the case of
the bidirectional coupling.

To prove the generality of our findings we have also studied
numerically two mutually coupled generators with tunnel
diodes~\footnote{
In this case Eq.~(\ref{eq:FunctRel}) should be
written as
$\mathbf{F}[\mathbf{x}(t),\mathbf{y}(t)]=0$.}.
In the
dimensionless form the dynamics of such generators is described by
the equations~\cite{Rosenblum:1997_SynhroHuman,alkor:TP2PS}
\begin{equation}
\begin{array}{l}
\dot
x_{1,2}=\omega^2_{1,2}[h(x_{1,2}-\varepsilon(y_{2,1}-y_{1,2}))+y_{1,2}-z_{1,2}],\\
\dot y_{1,2}=-x_{1,2}+\varepsilon(y_{2,1}-y_{1,2}),\\
\mu\dot z_{1,2}=x_{1,2}-f(z_{1,2}),
\end{array}
\label{eq:KPRs}
\end{equation}
where ${f(\xi)=-\xi+0.002\mathrm{sh}(5\xi-7.5)+2.9}$ is the
dimensionless characteristics of nonlinear converter, ${h=0.2}$,
${\mu=0.1}$, ${\omega_1=1.09}$, ${\omega_2=1.02}$ are the control
parameter values, $\varepsilon$ is the coupling parameter strength.
The indexes ``1'' and ``2'' correspond to the first and second
coupled systems, respectively. For such values of the control
parameters the threshold of the generalized synchronization regime
determined by the moment of the transition of the second positive
Lyapunov exponent in the field of the negative
values~\cite{GS_bidir_NDES2010,alkor:PierceBidir2011TPL} is around
$\varepsilon_{GS}\approx 0.08$.

As in the case of R\"ossler systems considered above we have chosen
the reference point $\mathbf{x}_0$ of chaotic attractor of the first
oscillator randomly and analyze the behavior of its nearest
neighbors $\mathbf{x}_i$ ($i=1,\dots,N$) and corresponding to them
vectors $\mathbf{y}_i$ and $\mathbf{z}_i$. The choice of the vector
basis $\mathbf{x}_{1-3}$ has been performed in the same way as in
the case considered above.

\begin{figure}[tb]
\centerline{\includegraphics*[scale=0.4]{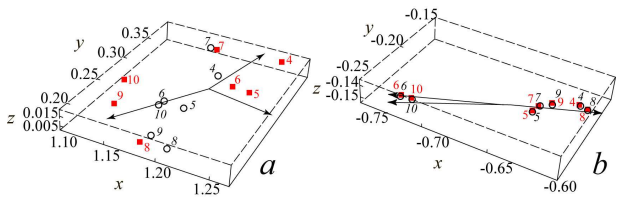}} \caption{(Color
online) The vectors $\mathbf{y}_i$ (\textcolor{red}{$\blacksquare$})
and $\mathbf{z}_i$ (\textcolor{black}{$\circ$}) of the second
generator with tunnel diode~(\ref{eq:KPRs}) for $\varepsilon=0.15$.
The numbers $i$ of the vectors are shown by the regular and italic
fonts, respectively. (\textit{a}) All neighbor vectors are used;
(\textit{b}) only vectors whose phase trajectories pass through the
phase tube with length $\tau=110$ are used} \label{fgr:VectorsTD}
\end{figure}

In Fig.~\ref{fgr:VectorsTD} the vectors $\mathbf{y}_i$
($\blacksquare$) and $\mathbf{z}_i$ (\textcolor{red}{$\circ$}) of
the second generator with tunnel diode~(\ref{eq:KPRs}) for the
coupling parameter strength $\varepsilon=0.15$ exceeding greatly the
threshold value of the generalized synchronization regime onset $\varepsilon_{GS}$ are
shown. Fig.~\ref{fgr:VectorsTD},\textit{a} corresponds to the case
when all neighbor vectors are used whereas in
Fig.~\ref{fgr:VectorsTD},\textit{b} only vectors whose phase
trajectories pass through the phase tube with length $\tau=110$ are
used. It is clearly seen that in the first case the vectors
$\mathbf{z}_i$ and $\mathbf{y}_i$ differ from each other
sufficiently testifying the failure of the presence of the
functional relation between the interacted system states.
But, conversely, for the phase tube with the length $\tau=110$ (Fig.~\ref{fgr:VectorsTD},\textit{b})
the calculated vectors $\mathbf{z}_i(t)$ are in the
excellent agreement with the vectors $\mathbf{y}_i(t)$ of the second
generator that confirms the results obtained above for
unidirectionally coupled R\"ossler systems. So, in the systems with
a mutual type of coupling the vector states of the interacting systems are related with each other by \emph{the functional}.

In conclusion, we have reported that the concept of generalized
synchronization \alkor{(except for the LS and CS regimes)} needs refining, since the state vectors of the
interacting chaotic systems are related with each other by the
\emph{functional}, but not the \emph{functional relation} as it was
assumed until now. \alkor{Although in the Report the systems with a small number of degrees of freedom have been considered, the developed formalism can be also extended to the systems with the infinite-dimensional phase space~\footnote{\alkor{In this case the system state is defined uniquely by the function (or vector-function) but not by the finite-dimensional vector as in the case of the system with small number of degrees of freedom.}}.}
Fortunately, this modification of the generalized
synchronization concept does not discard the majority of the
obtained hitherto results concerning GS. At the same time, this
refinement has a fundamental significance from the point of view of
the understanding of the core mechanisms of the considered phenomena
and is supposed to give a strong potential for new approaches and
applications dealing with the nonlinear systems.

\end{document}